\documentclass[pra,amssymb,amsmath,superscriptaddress,twocolumn,nofootinbib]{revtex4-1}

\usepackage{graphicx,color}
\usepackage[version=3]{mhchem}

\graphicspath{{pix/}}

\begin{document}

\title[]{Measurement of charge-exchange between Na and \ce{Ca+} in a hybrid trap}
\author{J.~M. Kwolek}
\email{jonathan.kwolek@uconn.edu}
\affiliation{Department of Physics, University of Connecticut, Storrs, Connecticut 06268}

\author{D.~S. Goodman}
\affiliation{Department of Chemistry and Physical Sciences, Quinnipiac University, Hamden, Connecticut 06518}
\affiliation{Department of Physics, University of Connecticut, Storrs, Connecticut 06268}

\author{B. Slayton}
\affiliation{Department of Sciences, Wentworth Institute of Technology, Boston, Massachusetts 02115}

\author{R. Bl\"umel}
\affiliation{Department of Physics, Wesleyan University, Middletown, Connecticut 06459}

\author{J.~E. Wells}
\affiliation{W. M. Keck Science Department of Claremont McKenna, Pitzer, and Scripps Colleges, Claremont, California 91711}
\affiliation{Department of Physics, University of Connecticut, Storrs, Connecticut 06268}

\author{F.~A. Narducci}
\affiliation{Department of Applied Physics, Naval Postgraduate School, Monterey, California 93943}

\author{W.~W. Smith}
\affiliation{Department of Physics, University of Connecticut, Storrs, Connecticut 06268}
\date{\today}

\begin{abstract}
We present measurements of the charge-exchange reaction rate between neutral sodium (Na) and ionized calcium (\ce{Ca+}) in a hybrid atom-ion trap, which is comprised of a Na magneto-optical trap concentric with a linear Paul trap. Once the Na and \ce{Ca+} are co-trapped, the reaction rate is measured by continuously quenching the reaction product \ce{Na+} from the ion trap, and then destructively measuring the decay of the remaining ion population. The reactants' electronic state and temperature are experimentally controlled, allowing us to determine the four individual reaction-rates between $\text{Na}[\text{S~or~P}]$ and $\text{Ca}^+[\text{S~or~D}]$ at different collision energies. With the exception of the largest reaction-rate channel ($\text{Na}[\text{S}]+\text{Ca}^+[\text{D}]$), our rates agree with classical Langevin rate limit. We have also found evidence of reactant collision-energy thresholds associated with two of the four entrance-channels.

\end{abstract}

\maketitle

\section{Introduction}

The long-range ion-neutral interaction arises from the attraction of the ion to the polarized neutral atom, and has the characteristic potential \mbox{$V\propto -1/R^4$} at a large distance $R$. Consequently, at ultracold temperatures, the ion-neutral elastic and charge-exchange cross sections can be millions of times larger than those of atom-atom collisions \cite{cote_ultracold_2000, cote_classical_2000}.

Cold to ultracold ion-neutral interactions were relatively unexplored until the creation of the hybrid trap \cite{smith_collisional_2003}, a device capable of co-trapping and cooling ions and neutral atoms \cite{smith_cold_2005}. Much early work with the hybrid traps focused on sympathetically cooling the translational motion of atomic ions and the internal internal degrees of freedom of molecular ions using the co-trapped neutral species \cite{dutta_cooling_2018,schowalter_blue-sky_2016,sivarajah_evidence_2012,goodman_ion-neutral-atom_2012,ravi_cooling_2012,zipkes_trapped_2010,rellergert_evidence_2013}.

Most successfully, the hybrid trap has been used to study ion-neutral chemical reactions at low temperature, where quantum effects are important. The hybrid trap allows for precise control over the reactants' electronic state and temperature \cite{ratschbacher_controlling_2012}, allowing both accurate and precise measurements of reaction rates and branching ratios. Ion-neutral quantum chemistry is of great interest to cosmology, since some of these reactions play a significant role in the formation of the universe \cite{smith_ion_1992,reddy_first_2010,stancil_radiative_1996} and interstellar medium \cite{smith_ion_1992}. Additionally, quantum chemistry offers the possibility of creating mesoscopic molecules \cite{cote_mesoscopic_2002} or trapped ultracold molecular ions \cite{hudson_method_2009,sullivan_trapping_2011,puri_synthesis_2017} that could be used for modern quantum-information applications. Furthermore, studying ion-neutral interactions expands our understanding of both the beneficial and potentially undesirable chemistry associated with candidates for trapped-ion atomic clocks and ion-based quantum information architectures \cite{idziaszek_quantum_2009,kollath_scanning_2007,zipkes_trapped_2010,ratschbacher_decoherence_2013}.

Measurements of ion-neutral reaction rates have both spurred efforts to calculate fully-quantal molecular potential curves and the associated reaction rates and allowed those calculations to be evaluated against experimental results. For example, in cosmological modeling, the variation in the predicted ion-neutral reaction-rates, like those of associative-detachment reactions, can lead to orders-of-magnitude discrepancies in protogalactic collapse models \cite{glover_cosmological_2006}, making the accuracy of computational predictions critical.

Hybrid traps have been used to measure charge exchange in many heteronuclear atomic systems, including $\text{Rb}+\text{Yb}^+$ \cite{lee_measurement_2013}, $\text{Rb}+\text{Ba}^+$ \cite{ravi_cooling_2012}, $\text{Rb}+\text{Sr}^+$ \cite{meir_dynamics_2016}, $\text{Rb}+\text{Ca}^+$ \cite{hall_ion-neutral_2013}, $\text{Rb}+\text{K}^+$ \cite{dutta_collisional_2017}, $\text{Rb}+\text{Cs}^+$ \cite{dutta_cooling_2018}, $\text{Li}+\text{Ca}^+$ \cite{haze_charge-exchange_2015}, $\text{Li}+\text{Yb}^+$ \cite{joger_observation_2017}, $\text{Ca}+\text{Yb}^+$ \cite{rellergert_measurement_2011}, $\text{Ca}+\text{Ba}^+$ \cite{sullivan_role_2012}, $\text{Na}+\text{Ca}^+$ \cite{smith_experiments_2014}, and $\text{Cs}+\text{Rb}^+$ \cite{dutta_collisional_2017}. Additionally, studies have been done in the molecular system $\text{Rb}+\text{N}_2^+$ \cite{hall_millikelvin_2012}, as well as the homonuclear systems $\text{Rb}+\text{Rb}^+$ \cite{lee_measurement_2013,ravi_cooling_2012}, $\text{Cs}+\text{Cs}^+$ \cite{dutta_cooling_2018}, $\text{Yb}+\text{Yb}^+$ \cite{grier_observation_2009}, and $\text{Na}+\text{Na}^+$ \cite{goodman_measurement_2015}.

Previously, our group reported evidence for a $\text{Na}+\text{Ca}^+$ charge-exchange reaction \cite{smith_experiments_2014}, which is predicted to be too small to observe for the A-X reaction channel \hbox{($\sim 10^{-16}\;\text{cm}^3/\text{s}$)} \cite{makarov_radiative_2003}. Here, we revisit the $\text{Na}+\text{Ca}^+$ experiment and expand upon that initial work discussed in Ref.~\cite{smith_experiments_2014}. Specifically, we measure the charge-exchange rate over a range of collision energies and we investigate individual reaction entrance channels by controlling the electronic state of the reactants. Also, we utilize our study of the sodium magneto-optical trap (MOT) excited-state fraction \cite{kwolek_model-independent_2018} to more accurately determine the quantum state of the neutral Na in our experiment, thus increasing accuracy and specificity of our rate-coefficient measurements.

The paper is organized as follows: In Sec.~\ref{sec:app}, we briefly describe the apparatus and its details relevant to this experiment. In Sec.~\ref{sec:expt}, we describe the model and measurement-methods used to experimentally determining the charge-exchange rate constant. In Sec.~\ref{sec:results}, we discuss the results of our measurements, which are compared with classical theoretical rate predictions. We conclude in Sec.~\ref{sec:concl}. 

\section{Apparatus}
\label{sec:app}
We will briefly discuss our hybrid trap apparatus \cite{goodman_measurement_2015,kwolek_model-independent_2018,wells_loading_2017} and highlight the aspects unique to this experiment.

\begin{figure}
\includegraphics[width=\linewidth,trim={1em 3em 5em 0}]{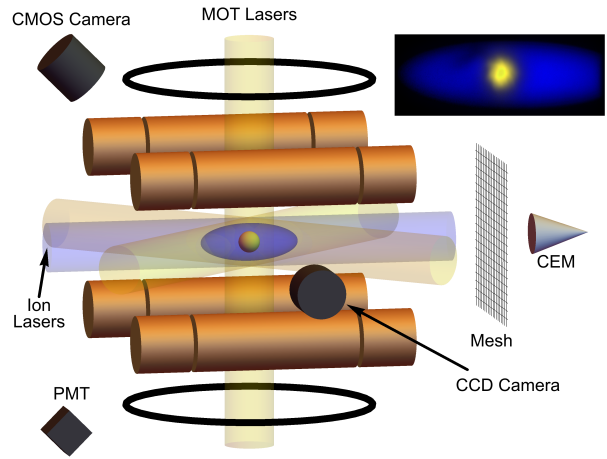}
\caption{(Color online)  The hybrid trap consists of a concentric MOT and linear Paul trap (LPT). The Na MOT consists of three pairs of orthogonal 589 nm laser beams and anti-Helmholtz coils. The LPT consists of four sets of segmented copper-rods. An rf voltage is applied to the center rods, while the end-segments are kept at a dc voltage. The \ce{Ca+} is created by exciting neutral background Ca vapor with 423 nm light, and photoionizing the excited Ca with broadband 375 nm light. Additional lasers include a 397 nm laser for Doppler cooling or shelving the \ce{Ca+} in the $\text{D}_{3/2}$ state and an 866 nm laser for repumping back into the $\text{P}_{1/2}$ state. Measurement of the ion population is performed by gating the end segments into a dipole configuration, so that the ions are ejected from the trap, traveling through the low-voltage mesh and into the channeltron-electron multiplier (CEM). Atom and ion fluorescence measurements can be performed with the CCD and CMOS cameras, or with a photomultiplier tube (PMT). The inset shows a false-colored image from the CCD camera of the co-trapped Na MOT (yellow) and \ce{Ca+} ion cloud (blue).}
\label{apparatus}
\end{figure}

Our hybrid trap consists of a concentric linear Paul trap (LPT) \cite{paul_electromagnetic_1990} and a Na magneto-optical trap (MOT) \cite{raab_trapping_1987}, housed within a vacuum chamber held at a pressure of $\sim 10^{-10}$ Torr. A diagram of our trap can be seen in Fig.~\ref{apparatus}.

The Na MOT consists of a pair of anti-Helmholtz electromagnets external to the vacuum chamber and three pairs of oppositely circularly-polarized counter-propagating laser beams along each orthogonal dimension. A continuous-wave Toptica MOT laser is tuned to the Na $S_{1/2}\to P_{3/2}$ D2 manifold, near 589 nm to create the MOT beams. The MOT is vapor-loaded from a hot neutral Na source. The atoms can be trapped using two different hyperfine transitions, the so-called type-I and type-II MOTs \cite{jarvis_blue-detuned_2018,oien_cooling_1997}. The type-I MOT is made by tuning the MOT laser to the $F=2\to F'=3$ transition. The type-I MOT used here typically has a temperature of $\approx 300\;\mu\text{K}$, atom density of $\sim 10^{10}\;\text{cm}^{-3}$, and population of $\sim 10^6\;\text{atoms}$. The type-II MOT is made by tuning the MOT laser to the $F=1\to F'=0$ transition. The type-II MOT used here typically has a temperature of $\approx 4\;\text{mK}$, atom density of $\sim 10^{9}\;\text{cm}^{-3}$, and population of $\sim 10^7\;\text{atoms}$. In each case, the 589 nm laser is modulated by an electro-optic modulator (EOM), which creates sidebands. These sidebands can be used to either trap atoms or to repump them out of the dark states. Our experiment primarily used the type-II MOT, which uses the carrier as the repump laser and a sideband as the cooling laser. We can modulate the control voltage on the EOM to shutter the MOT on and off by optically pumping the atoms into a dark hyperfine state, which takes $\approx 10\;\mu$s. Because a MOT is made with near-resonance lasers, we can use the intensity and detuning of these lasers to control the steady-state population of the neutral atoms in the excited state. We have quantified the MOT's excited-state fractional-population $f_e$ over our entire range of MOT operating parameters \cite{kwolek_model-independent_2018}.

The LPT consists of four segmented rods, and is responsible for trapping the \ce{Ca+} ions. When trapping, we apply a 780 kHz, 60-V amplitude radio-frequency (rf) signal to each of the central rods, confining the ions radially. A 30-V static potential is applied to the outer segments, confining the ions axially. One diagonal pair of central rods is 180 degrees out-of-phase with the other, creating the time-averaged harmonic trapping potential \cite{paul_electromagnetic_1990}. Neutral Ca is vaporized from a hot Ca source. Ions are loaded into the trap by two cw lasers in the center of the trapping region; a Toptica 423 nm diode laser resonantly excites the \ce{^1 S_0} to \ce{^1 P_1} transition in the Ca vapor and an RGBLase broadband 375 nm diode laser photoionizes the excited Ca.

We can control the optically addressable electronic state of our \ce{Ca+} ions with Toptica 397 nm and 866 nm diode lasers. We excite the \ce{Ca+} on the transition between the $\text{S}_{1/2}$ and $\text{P}_{1/2}$ with the 397 nm light. Ions in the $\text{P}_{1/2}$ state can decay into the $\text{D}_{3/2}$ state, which is metastable, with a \hbox{$\sim 1$ s} lifetime \cite{NIST_ASD}. When laser cooling on the \ce{Ca+} $\text{S}_{1/2}\to\text{P}_{1/2}$ transition, we can repump on the $\text{D}_{3/2}\to\text{P}_{1/2}$ transition at \mbox{866 nm}. Alternatively, we can optically pump the \ce{Ca+} into the metastable $\text{D}_{3/2}$ state by withholding the 866 nm radiation.

During the charge-exchange process, \ce{Na+} product ions are created in the trapping region. To ensure our extracted ion signal is entirely caused by \ce{Ca+}, we continuously quench the \ce{Na+} by applying a small oscillating voltage \mbox{($1$ V, $\approx 280\;\text{kHz}$)} to the rods resonant with the second harmonic of the trapped \ce{Na+} secular frequency, a process called mass-selective resonance quenching (MSRQ) \cite{sivarajah_off-resonance_2013}. With the MSRQ field on, the \ce{Na+} ions are expelled from the trap within a few secular oscillation cycles \cite{goodman_ion-neutral-atom_2012,sivarajah_off-resonance_2013}. We measure our total ion population after a variable co-trapping time by gating the end-segments into a dipole configuration and measuring the ion current incident on a \textsc{megaspiraltron} Channeltron electron multiplier (CEM). We can determine the time evolution of the trapped ion population by repeatedly loading and extracting the ions for different co-trapping intervals.

The photoionization threshold for excited $\text{Na}[3^2\text{P}_{3/2}]$ is $\approx 409$ nm \cite{NIST_ASD}. Consequently, the 397 nm light used for cooling and shelving \ce{Ca+} efficiently photoionizes the excited Na in our MOT \cite{wells_loading_2017}. The continuous quenching of photoionized \ce{Na+} via MSRQ sympathetically heats the co-trapped \ce{Ca+} ions. To eliminate the photoionization process, we asynchronously chop the 397 nm beam with a mechanical chopper-wheel and modulate the MOT's 589 nm beam with the EOM. We run our chop with a frequency of 5 kHz and a MOT duty-cycle of 45\%, which results in a 99.96\% reduction in photoionization rate on the MOT by the 397 nm laser. The chop period is much less than the D-state relaxation time, which means that any ions shelved in the D state will remain in that state during the 397 nm laser-off phase of the chop cycle.

Last, we can measure ion- or atom-fluorescence with our photomultiplier tube (PMT) or cameras. To image both the atomic and ionic clouds directly, we aim two cameras (Mightex 1.3 MP monochrome CCD and a Thorlabs 1.3 MP color CMOS) into the chamber from orthogonal directions. Each camera is sensitive to both 397 nm and 589 nm light.

\section{Experiment}
\label{sec:expt}
In Sec. A we discuss the decay model, including its limitations. We will discuss the decay measurement itself in Sec. B. Next, in Sec. C, we explain how the atom and ion temperatures are determined, and how these results are used in our analysis of the decay results. Lastly, in Sec. D, we discuss the measurement of the ion-cloud size, which we use to quantify the spatial overlap of our atom and ion distributions.
\subsection{The Decay Model}
The loss of ions from the trap can be modeled by the linear differential rate equation, 
\begin{equation}
\frac{\text{d}N_I}{\text{d}t}=-(\gamma_\mathrm{ia}+\gamma_b)N_I, \label{eq:diffeq}
\end{equation}
where $N_I$ is the total ion number, and $\gamma_\mathrm{ia}$ and $\gamma_b$ are the loss rates due to charge-exchange (assisted by MSRQ of the \ce{Na+} product) and non-charge-exchange loss, respectively. The non-charge-exchange term comes from the inherent ion-trap loss-mechanisms such as excess micromotion \cite{berkeland_minimization_1998} or rf heating \cite{tarnas_universal_2013}. The trap's characteristic lifetime is $\tau = 1/\gamma_b$.

The solution to Eq.~\eqref{eq:diffeq} is,
\begin{equation}
N_I(t)=Ae^{-(\gamma_\mathrm{ia}+\gamma_b)t}, \label{eq:exp}
\end{equation}
with a total rate of $\gamma_\mathrm{ia}+\gamma_b$. We normalize our decays to the first measured data point, such that $A=1$. The charge-exchange rate-coefficient is defined as
\begin{equation}
k_\mathrm{ia}=\frac{\gamma_\mathrm{ia}}{\langle n\rangle}, \label{eq:kia}
\end{equation}
where $\langle n\rangle$ is a measure of the average atom-overlap density with the ion distribution. Assuming a Gaussian distribution of both atoms and ions \cite{rellergert_measurement_2011,sullivan_role_2012,grier_observation_2009,goodman_measurement_2015}, we can model their overlap as
\begin{equation}
\langle n \rangle=N_a\prod_{i=\{x,y,z\}}\left(\int_{-\infty}^\infty\frac{e^{-(x_i-x_{0,i})^2/r_{a,i}^2}}{\sqrt{\pi} r_{a,i}}\frac{e^{-x_i^2/r_{I,i}^2}}{\sqrt{\pi} r_{I,i}}\;dx_i\right). \label{eq:overlap}
\end{equation}
Here, $r_{a,i}$ and $r_{I,i}$ represent the radius of the $i$th dimension of the atom and ion distribution, respectively. Additionally, $x_{0,i}$ is the offset of the atom distribution from the center of the ion distribution and $N_a$ represents the total atom number. Before taking ion decay data, we first image the MOT and the laser-cooled \ce{Ca+} ion-cloud separately, to ensure maximum concentricity of the two clouds, such that  $x_{0,i} \approx 0$, where $x_{0,i}$ is no larger than 5\% of a MOT radius for $i =$ 1, 2, and 3.

\subsection{Decay Measurement}
\label{subsec:decay}
\begin{figure}[ht!]
\includegraphics[width=\linewidth]{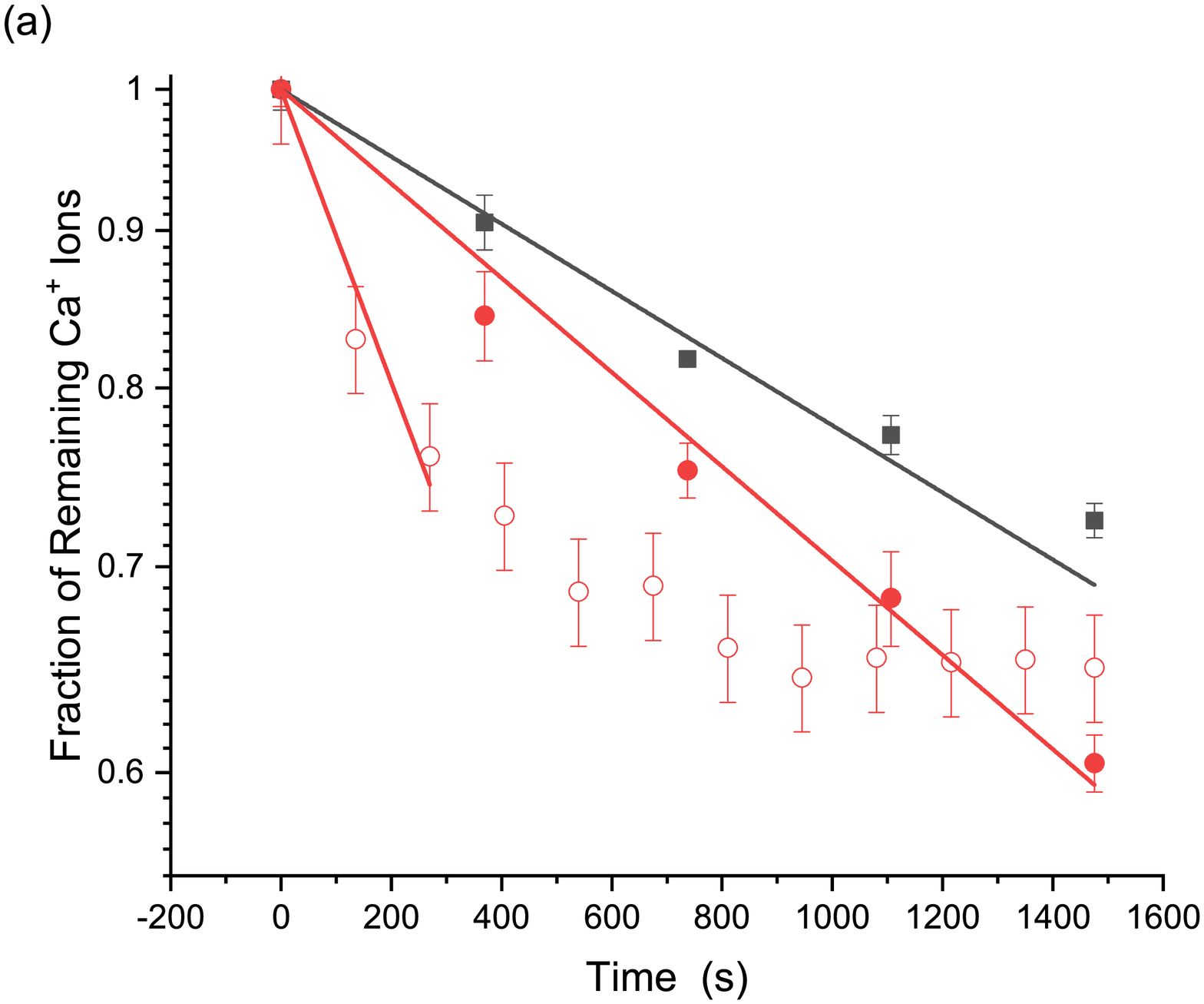}
\includegraphics[width=\linewidth]{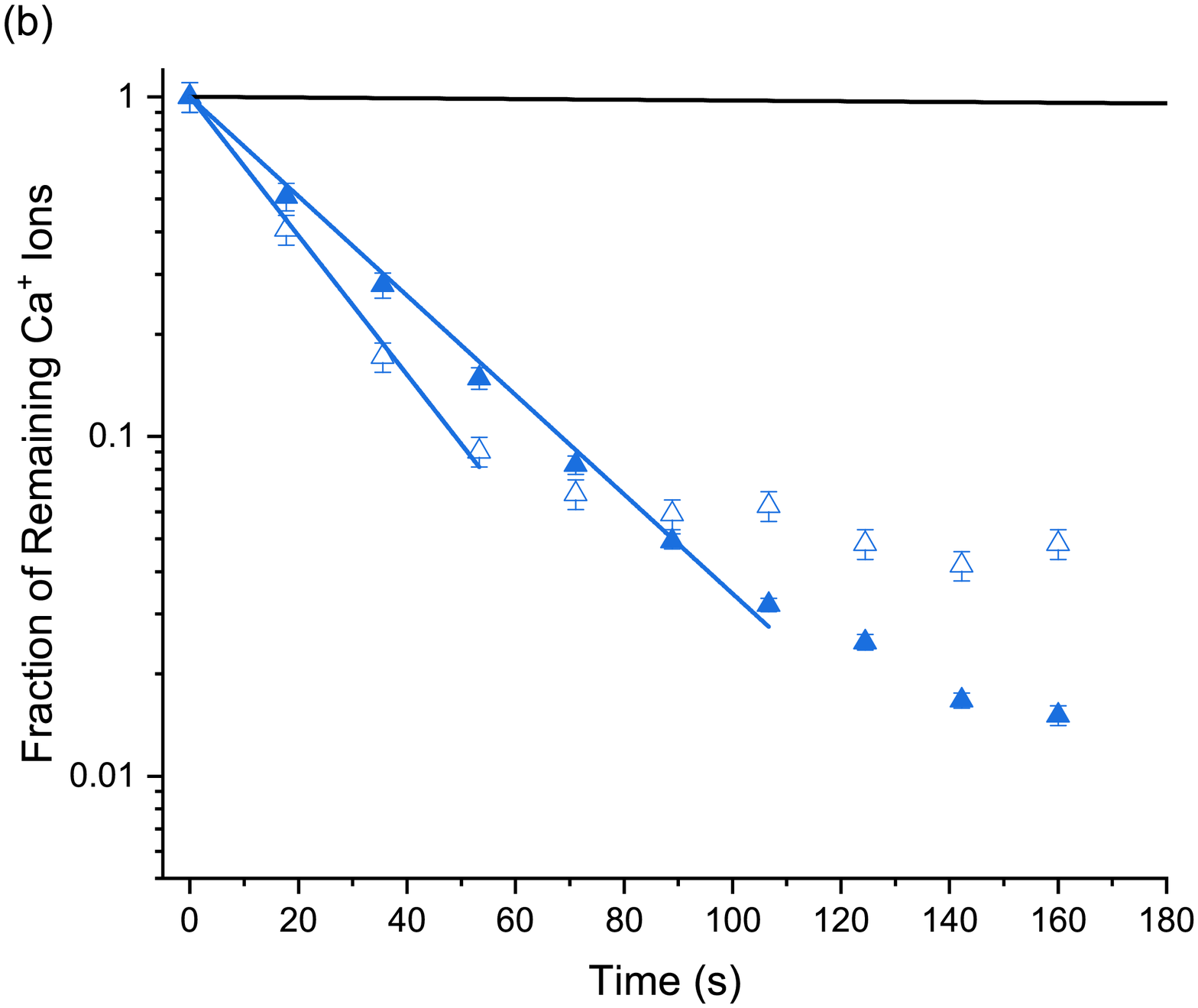}
\caption{\label{decays} (Color Online) The Ca$^+$ trapped-ion population decay from Ca$^+$[S] (a) and Ca$^+$[D] (b) as a function of trapping time on a log-linear plot. The background loss (black squares) is shown in plot (a) and the corresponding fit (solid black line) is shown in both plots (a) and (b). When we introduce a MOT into the hybrid trap, there is an additional loss rate from the LPT due to charge-exchange reactions between the ionized \ce{Ca+} and Na. By changing the electronic-state of the \ce{Ca+}, we observe drastically different reaction rates between the \ce{Ca+} S-state (red unfilled circles) and D-state (blue unfilled triangles) entrance channels. The decay curves are not purely exponential, a behavior that is correlated with a change in ion-cloud temperature and may be indicative of a reaction collision-energy threshold. After heating the sample (filled markers), the initial single-exponential behavior is observed over a larger percentage of the decay.  We fit the curves to the exponential solution of Eq.~\eqref{eq:diffeq} in the initial region of constant ion-cloud temperature, as seen in Fig.~\eqref{iontemp}. The error bars are statistical, calculated from repeated measurements.}
\end{figure}

To determine $k_\mathrm{ia}$, we separately measure $\gamma_\mathrm{ia}$, $\gamma_{b}$, and $\langle n\rangle$. First, we measure the background loss rate $\gamma_b$ by measuring the trapped \ce{Ca+} decay with the MSRQ fields turned on (resonant with \ce{Na+}), but the MOT turned off. We can confirm by running the same test with the MSRQ off that the MSRQ field has no observable effect on the \ce{Ca+} lifetime. We typically find the trap's characteristic lifetime without charge exchange to be \mbox{$\tau \approx 4000$ s}, as seen in Fig.~\ref{decays}.  Second, we repeat this measurement, but now with the MOT turned on and observe a faster decay rate, $\gamma_\mathrm{ia}+\gamma_b$.

In the absence of \ce{Ca+} excitation lasers, the ions remain in the S-state and react with a mixture of excited and ground-state Na. Alternatively, we can use the 397 nm laser to shelve the \ce{Ca+} within the D-state and observe this reaction with the same mixture of excited and ground-state Na. The results of a typical charge-exchange measurement with S and D-state \ce{Ca+} can be seen in Fig.~\ref{decays}, which shows that the D-state reaction is significantly faster than that of the S-state.

Equation~\eqref{eq:exp} can be applied to the initial part of the decay curve, but eventually breaks down. There is a clear deviation from simple exponential decay in the reaction decay curves; the onset of the deviation depends on a change in temperature of the ion cloud, as discussed in \mbox{Sec.~\ref{subsec:temp}} and shown in Fig.~\ref{iontemp}. We found that the temperature of the ion cloud decreases after the point at which the decay data deviate from a simple exponential.  However, a temperature decrease should increase the overlap density $\langle n\rangle$, since the cloud temperature is correlated with its \cite{berkeland_minimization_1998}. This means that a decrease in temperature would make the decay steeper (faster), not shallower as seen in Fig.~\ref{decays}. The slowing of the decay rate at the breakaway point suggests that the now colder ion-neutral reactants essentially stop reacting, possibly due indicating that there is a minimum energy required for the reaction to progress. This hypothesis is further supported by the observation that the single-exponential decay model fits the data for longer co-trapping times when the ion-cloud temperature is increased, as seen by comparing the curves with filled markers to those with unfilled markers in Fig.~\ref{decays}. Presumably, the hotter \ce{Ca+} population has a larger fraction of ions whose energy is above the collision-energy threshold. 

To rule out any systematic errors with our destructive CEM ion detection scheme, we independently measured the \ce{Ca+} population decay using our PMT. The measurement procedure was the same, except just before the ions are extracted, the \ce{Ca+} cloud was illuminated with resonant 397 nm radiation and the resulting \ce{Ca+} fluorescence was measured with the PMT.  We found that the decay in the brightness of the \ce{Ca+} cloud followed the same curve as that of the CEM data shown in Fig.~\ref{decays}. Additionally, to ensure that the only ions trapped were \ce{Ca+} (and not molecular \ce{NaCa+}), we repeated the experiment, but with a sudden switch of the MSRQ frequency to that of the \ce{Ca+} secular frequency right before detecting the ion-population with the CEM. Consequently, any trapped \ce{Ca+} was quickly ejected before detection and no ion-signal was observed, suggesting that the only ions trapped were indeed \ce{Ca+}.  

\subsection{Ion and Atom Temperature}
\label{subsec:temp}

\begin{figure}
\includegraphics[width=\linewidth]{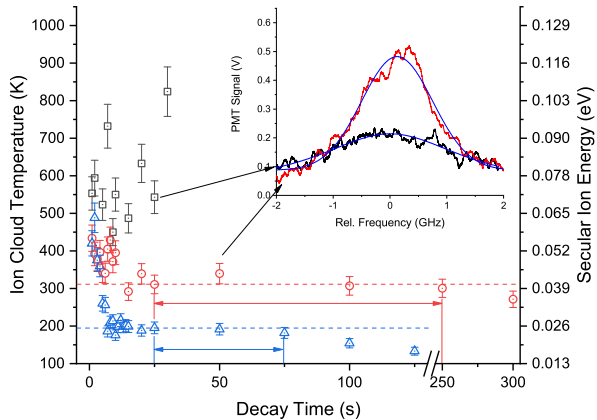}
\caption{\label{iontemp} (Color online) The temperature of the ion cloud changes as a function of time in the trap. The temperature of the ion cloud is determined by fitting a fluorescence measurement of a scanning 397 nm laser over an electronic transition in \ce{Ca+} (inset). This measurement is fit to a Voigt profile, to give an estimate of the temperature distribution in the cloud. Without a MOT present (black squares), the trap heats due to inherent LPT heating-mechanisms until the ion cloud is too dilute to make an accurate measurement with our PMT. With a MOT present, the ion cloud is sympathetically cooled to different steady-states, depending on the ion target state (S-state, red circles; D-state, blue triangles). The arrows near the $x$-axis show the range of times used for fitting decay curves, like those shown in Fig.~\ref{decays}. The horizontal dashed lines are drawn to illustrate that the temperature does not measurably change during our range of decay-times chosen (horizontal arrows) for fitting \ce{Ca+} decay curves to Eq.~\eqref{eq:exp}. Temperature is converted to secular-energy by Eq.~\eqref{secen}. Each error in temperature is statistical, calculated from repeated measurements.}
\end{figure}

To better understand the decay curve's deviation from Eq.~\eqref{eq:exp}, the ion and atom cloud temperatures needed to be determined. The ion cloud temperature can be measured via ion fluorescence using a fast ($\sim 1\;\text{kHz}$) and weak ($<1\;\text{mW}$) \hbox{397 nm} laser scan, so as to measure the ion cloud's temperature without simultaneously cooling the cloud. We concurrently illuminate the cloud with the \hbox{866 nm} laser, in order to prevent optical pumping into the dark D-state. We fit this scan to a Voigt profile, with a power-broadened Lorentzian linewidth \cite{kielpinski_laser_2006}; examples of these temperature measurements are shown in the inset of Fig.~\ref{iontemp}. By performing the temperature scan periodically throughout the co-trapping interval, we can better understand the temperature evolution during the time interval during which the reactions occur.

For comparison with potential theoretical predictions, we are primarily interested in the ion-atom collision energy $\frac{1}{2} \mu v^2$, where $\mu$ is the two-body reduced mass and $v$ is the relative ion-atom velocity. We can convert the ion temperature into an average energy by considering the secular and micromotion separately. Movement along the axial dimension of the trap will only consist of secular motion \cite{berkeland_minimization_1998}. According to the equipartition theorem, the total secular energy $E_s$ in the trap is 
\begin{equation}
E_{s}=\sum_i E_{s,i}=3 E_{s,z}=\frac{3}{2} k_B T,\label{secen}
\end{equation}
where $k_B$ is the Boltzmann constant, $T$ is the axially-measured temperature, and $E_{s,i}$ is the secular energy in the $i$th dimension. The secular energy in each dimension is the same, which is equivalent to the energy measured in the $z$-dimension, $E_{s,z}$. The micromotion energy will add to the secular energy in each radial trapping dimension \cite{berkeland_minimization_1998}, where the average ion energy in each dimension is
\begin{equation}
\langle E_i\rangle=E_{s,i}\left(1+\frac{q_i^2}{q_i^2+2a_i}\right).
\label{eq:E}
\end{equation}
Here, $q_i$ and $a_i$ are the so-called stability parameters for the Mathieu equation, which is the equation of motion for a single ion or dilute weakly-coupled cloud of ions trapped near the center of the trap. The first term in Eq.~\eqref{eq:E} is due to the secular energy and the second is due to the micromotion energy. Summing over all degrees of freedom, the average energy per ion time-averaged over the secular period is
\begin{equation}
\langle E\rangle=\frac{1}{2}k_B T\left(3+\frac{2 q^2}{q^2+2a}\right).\label{temp}
\end{equation}
In Eq.~\eqref{temp}, we used the fact that for a linear Paul trap $q_1 = q_2 \equiv q$ and $q_3 = 0$. Consequently, only $a_1 = a_2 \equiv a$ contribute to the average energy calculation. Typically in our experiment, $q\approx0.33$ and $a\approx0.004$, so the micromotion energy is $\approx 93 \%$ of the secular energy in each radial dimension, according to Eq.~\eqref{eq:E}.

We found that co-trapping the ion cloud and MOT reduces the ion energy (as seen in Fig.~\ref{iontemp}), presumably due to atom-ion sympathetic cooling of the non-laser-cooled \ce{Ca+} via elastic scattering with the laser-cooled Na \cite{sivarajah_evidence_2012,goodman_ion-neutral-atom_2012}. On the time scales set by the charge-exchange reactions the sympathetic cooling is effectively instantaneous. The steady-state temperature of the sympathetically cooled ion cloud can be increased above that from Eq.~\eqref{temp} by introducing excess micromotion \cite{berkeland_minimization_1998}, allowing us to explore different ranges of collision energy. The temperatures reported in Fig.~\ref{iontemp} correspond to zero excess micromotion, and thus the lowest achievable \ce{Ca+} temperatures in this experiment.

We found that the D-state ions were cooled more effectively than the S-state ions, which could be an indication of a cooling enhancement from a barrier in the molecular curve for the D-state entrance channel. Ignoring tunneling effects, a barrier in the entrance channel could increase the elastic cross section for collision energies less than the barrier height by increasing the effective interaction radius.

To ensure the ion-cloud size remains constant throughout the reaction, we are careful to fit our rate measurements within the region of post-cooling stability, as indicated in Fig.~\ref{iontemp}. Additionally, in order to compare with theoretical calculations of each $k_\mathrm{ia}$, which are typically averaged over a Maxwell-Boltzmann distribution of energies, we again must only fit our exponential decay during times when the temperature is not changing.

To measure the energy distribution of the MOT atoms, we use the fast EOM modulation to turn the MOT cooling force off for a variable amount of time, during which the MOT undergoes ballistic expansion. The change in size during this expansion can be used to determine the temperature of the MOT \cite{williams_characteristics_2017,lett_observation_1988}. We found that the typical temperature of the MOT to be $T= 300(50)\;\mu\text{K}$ and $T= 4(1)\;\text{mK}$ for the type-I and type-II MOTs, respectively. The MOT is many orders of magnitude colder than the ion cloud, so the ion energy distribution is approximately equivalent to the ion-atom relative collision-energy distribution.

\subsection{Ion-Cloud Size} 
With a value of $\gamma_\mathrm{ia}$, we can now determine $k_\mathrm{ia}$ by measuring the effective overlap volume, $\langle n\rangle$. First, we determine the total number of atoms $N_a$ and the MOT's spatial distribution by imaging the MOT with our cameras. Previously \cite{kwolek_model-independent_2018}, our group demonstrated that we can directly measure $f_e$, and thus $N_a$ from the MOT fluorescence.

When the MOT is not chopped, as in the case of the \ce{Ca+} S-state measurements, its size is determined by directly fitting the MOT image to a 2D Gaussian. However, special care is taken to determine the MOT's size while it is chopped, as in the \ce{Ca+} D-state measurements. While chopped, the MOT undergoes ballistic expansion during the off-phase of the cycle. Therefore, the time-averaged size of the MOT is approximately the average of the MOT's smallest size (bright-size) and largest size (dark-size). The ballistic expansions data used in determining the temperature of the MOT can also be used here to predict the MOT's dark-size at the end of the chopped off-phase. 

\begin{figure}
\includegraphics[width=\linewidth]{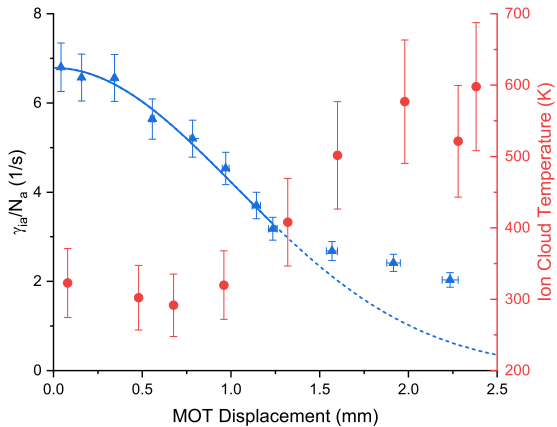}
\caption{\label{ionsize} (Color online) The atom-number normalized charge-exchange rate (blue solid triangles) changes as a function of MOT displacement in the radial dimension of the LPT. The change in the charge-exchange rate can be modeled by Eq.~\eqref{eq:concentricity} (solid and dashed blue line). In an LPT, ion temperature is correlated with size, so we only fit the data (solid blue line) where the temperature of the ion cloud (red circles) remains constant. Beyond the region of constant temperature (MOT displacement $>1.3$ mm), the concentricity function no longer fits. The extrapolation of the fit into this region is shown with a dashed blue line. The size of the cloud increases with the increase in temperature, so as expected, the experimental data falls above the dashed line. The error was determined for $\gamma_{ia}$ via a fit to the decay, and for $N_a$ by analyzing the systematic errors arising from the detection system \cite{kwolek_model-independent_2018}. The ion-cloud temperature measurement error is calculated by statistical averaging over many trials.}
\end{figure}

The ion cloud cannot be imaged in the same way that the MOT was imaged, since the target \ce{Ca+} states are the S-state and D-state, not the fluorescing P-state. Depending on the laser detuning, illuminating the dark cloud on the S to P transition would either result in Doppler cooling or heating, changing the size of the cloud. Instead, we can measure each cloud radius $r_{I,i}$ for the ion distribution by probing the ion cloud with the MOT itself. The presence of the MOT, too, influences the ion temperature. However, as explained below, its effect is understood and controlled. Assuming that the trap is radially symmetric about the extraction axis, we need only measure the axial and radial distributions. The distribution of the ion cloud is monitored by observing the change in the charge-exchange reaction rate as a function of MOT offset from the center of the ion-trapping region \cite{goodman_measurement_2015}. Evaluating Eq.~\eqref{eq:overlap} gives
\begin{equation}
\langle n\rangle=\frac{N_a C}{V_\mathrm{ia}} \label{eq:n},
\end{equation}
where we define the dimensionless concentricity function as
\begin{equation}
C=\prod_{i=\{x,y,z\}}e^{-x_{0,i}^2/(r_{a,i}^2+r_{I,i}^2)},
\end{equation}
and overlap volume
\begin{equation}
V_\mathrm{ia}=\pi^{3/2}\prod_{i=\{x,y,z\}}\sqrt{r_{a,i}^2+r_{I,i}^2}.
\end{equation} 
Now, we can substitute Eq.~\eqref{eq:n} in Eq.~\eqref{eq:kia} and solve for
\begin{equation}
\frac{\gamma_\mathrm{ia}}{N_a} = \frac{k_\mathrm{ia} C}{V_\mathrm{ia}}.
\label{eq:giaN}
\end{equation}

To determine the ion-cloud size, we translate the MOT using electromagnetic shim-coils along the radial ($x,y$) or axial ($z$) dimension and measure the reduction in the ion-atom decay rate as the overlap between the clouds is reduced. For a given dimension, almost all parameters in Eq.~\eqref{eq:giaN} can be grouped into a single constant, such that we only have one fitting parameter. For example, if the MOT is translated along the radial dimension $i=x$ where $x_{0,y}=x_{0,z}=0$, Eq.~\eqref{eq:giaN} reduces to
\begin{equation}
\frac{\gamma_\mathrm{ia}}{N_a}=k_0 e^{-x_{0,x}^2/(r_{a,x}^2+r_{I,x}^2)},\label{eq:concentricity}
\end{equation}
where the remaining terms are grouped into
\begin{equation}
k_0=\frac{k_\mathrm{ia}}{\sqrt{\left(r_{a,y}^2+r_{I,y}^2\right)\left(r_{a,z}^2+r_{I,z}^2\right)}}.
\end{equation}
We do not group $N_a$ into this set of constants so that we can re-measure and adjust for small variations in atom number for each data point. Thus, we fit the normalized rate $\gamma_\mathrm{ia}/N_a$ vs.~MOT displacement, as seen in Fig.~\ref{ionsize}. We have assumed that $k_\mathrm{ia}$ remains constant as the MOT is translated and we have directly confirmed with our cameras that the time-averaged size of the MOT remains constant as the MOT is translated.

However, the sympathetic cooling becomes less effective as the MOT is offset from the center of the LPT \cite{goodman_ion-neutral-atom_2012}. A measurement of the ion temperature reveals the regime in which we can accurately fit our MOT offset measurement, since temperature and size are correlated in an LPT. For small offsets (less than 2 MOT radii), the temperature remains constant, as seen in Fig.~\ref{ionsize}. However, once the displacement is too great, the temperature begins to rise, systematically causing the size of the cloud to increase. Fitting the temperature-stable region determined in measurements similar to those shown in Fig.~\ref{ionsize} to Eq.~\eqref{eq:concentricity}, we see that our rate data agree well with our model, except where it is expected to systematically disagree (after the temperature starts to rise due to extreme MOT offset). Because all of our charge-exchange measurements were conducted with the MOT concentric with the ion cloud, our fit of the size in the temperature-stable region will be equivalent to the size during the decay measurement. We determine the error in the ion-cloud size by analyzing the error in the fit of the Gaussian over the region of unchanging temperature. In this determination, we also account for the uncertainty in the MOT size. 

The cooling rate of the ion cloud depends on both the target state of the $\textrm{Ca}^+$ and the density of the MOT. To adjust for changes in temperature and size vs. target state, we separately measure the temperature and size for each combination of reactants.

\section{Rate-Measurement Results}
\label{sec:results}
The atoms within the MOT are in a mixture of excited Na[3P] and ground state Na[3S] atoms. Thus, any measurement of charge exchange between the ion cloud and the MOT measures at least two reaction channels simultaneously, and our $\gamma_\mathrm{ia}$ result is actually a weighted average of rates from the excited and ground states of Na. We can distinguish between the ground and excited-state sodium  contributions to the charge-exchange rates as $k_{ia}^{(e)}$ and $k_{ia}^{(g)}$ respectively. With these distinctions, we can rewrite Eq.~\eqref{eq:kia} in terms of the MOT excited state fraction, $f_e$, as
\begin{align}
\frac{\gamma_\mathrm{ia}}{\langle n\rangle}&=(1-f_e)k_\mathrm{ia}^{(g)}+f_e k_\mathrm{ia}^{(e)}\nonumber \\
&=k_\mathrm{ia}^{(g)}+f_e(k_\mathrm{ia}^{(e)}-k_\mathrm{ia}^{(g)}).\label{eq:linfit}
\end{align}
In other words, when $f_e=0$, $\gamma_\mathrm{ia}\langle n\rangle=k_\mathrm{ia}^{(g)}$ and when $f_e=1$, $\gamma_\mathrm{ia}\langle n\rangle=k_\mathrm{ia}^{(e)}$. The MOT will typically have an $f_e\in[0.02,0.2]$. We can extrapolate a rate for \ce{Ca+} on both excited and ground-state Na from a linear fit to a plot of $\gamma_\mathrm{ia}/\langle n\rangle$ vs $f_e$.

\begin{figure}
\includegraphics[width=\linewidth]{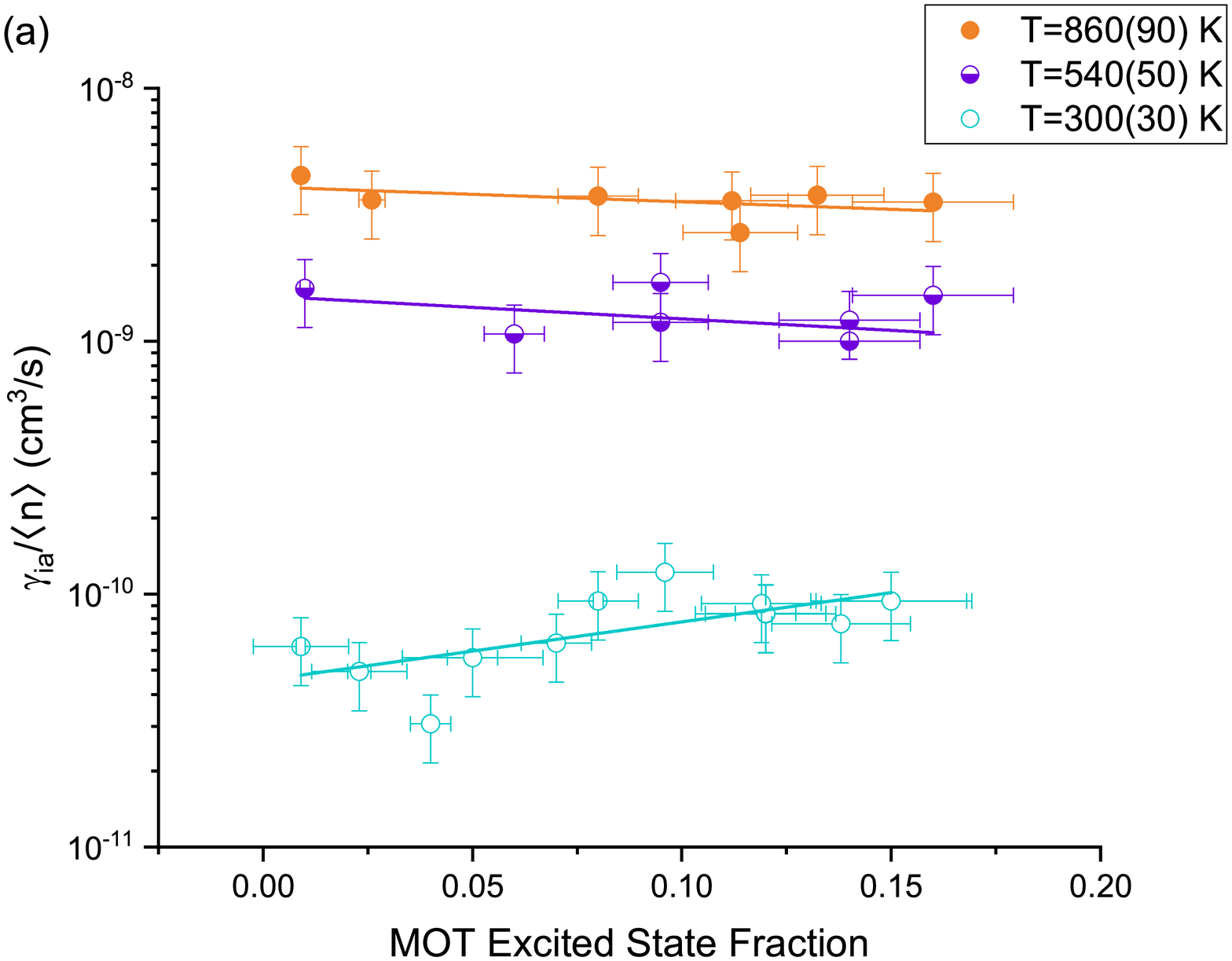}
\includegraphics[width=\linewidth]{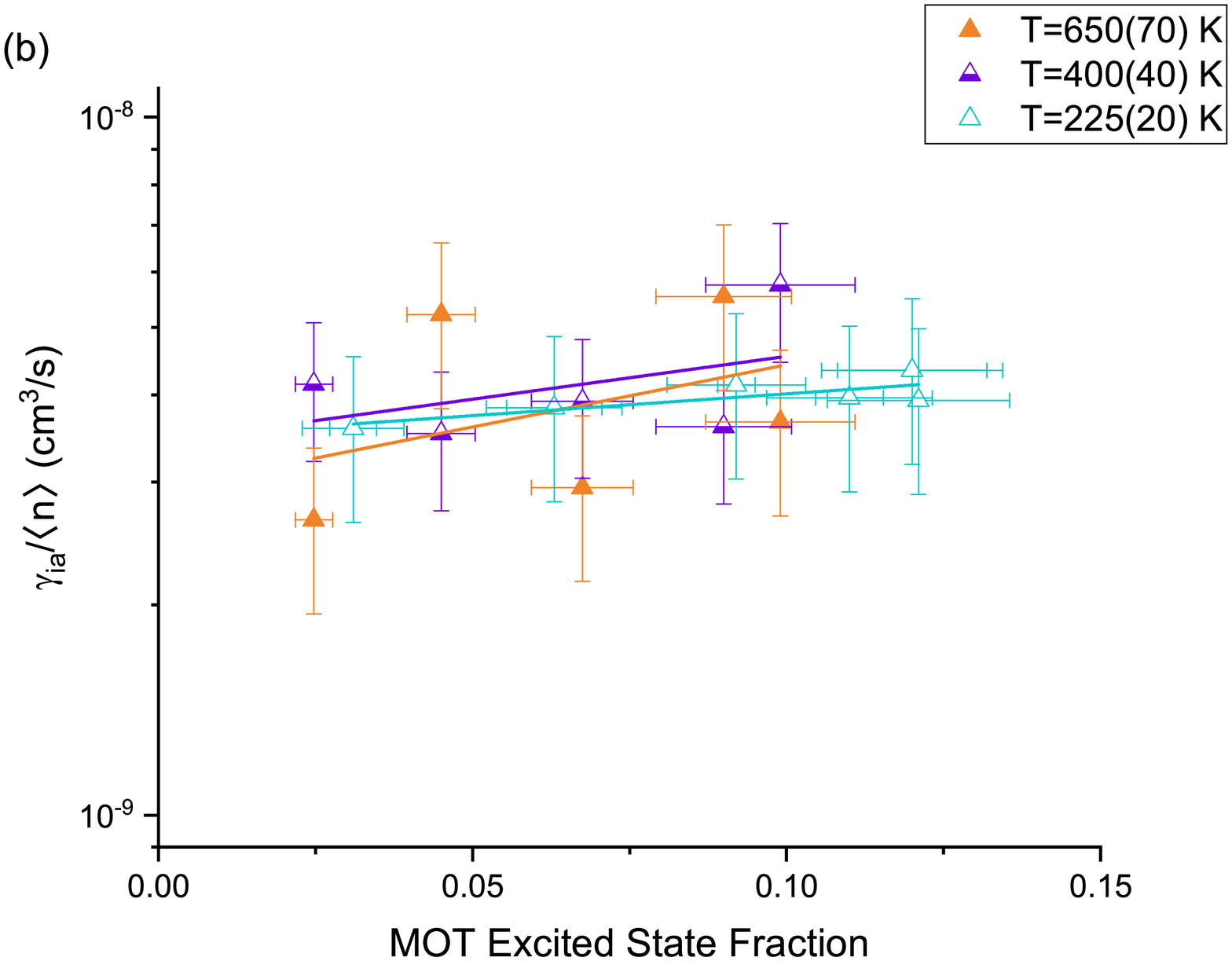}
\caption{\label{ratefe} (Color online) The rate $\gamma_\mathrm{ia}/\langle n\rangle$ is plotted as a function of MOT excited-state fraction for S-state (a) and D-state (b) \ce{Ca+}. A linear fit to the data is used to extract the reaction rates of the ground- and excited-state Na atoms, according to Eq.~\eqref{eq:linfit}. As the temperature increases, there is an increase in reaction rate for the $\text{Na}[\text{S}]+\text{Ca}^+[\text{S}]$ reaction channel. The \ce{Ca+} D-state reaction is fairly independent of $f_e$ or temperature. The resulting rates for the individual entrance channels are plotted in Fig.~\ref{kvt}. The error in $f_e$ is discussed extensively in Ref. \cite{kwolek_model-independent_2018}. Error in the rate $\gamma_{ia}/\langle n\rangle$ is calculated by propagating the fit-error of $\gamma_{ia}$ and the uncertainty from the overlap measurements.}
\end{figure}

\begin{figure}
\includegraphics[width=\linewidth]{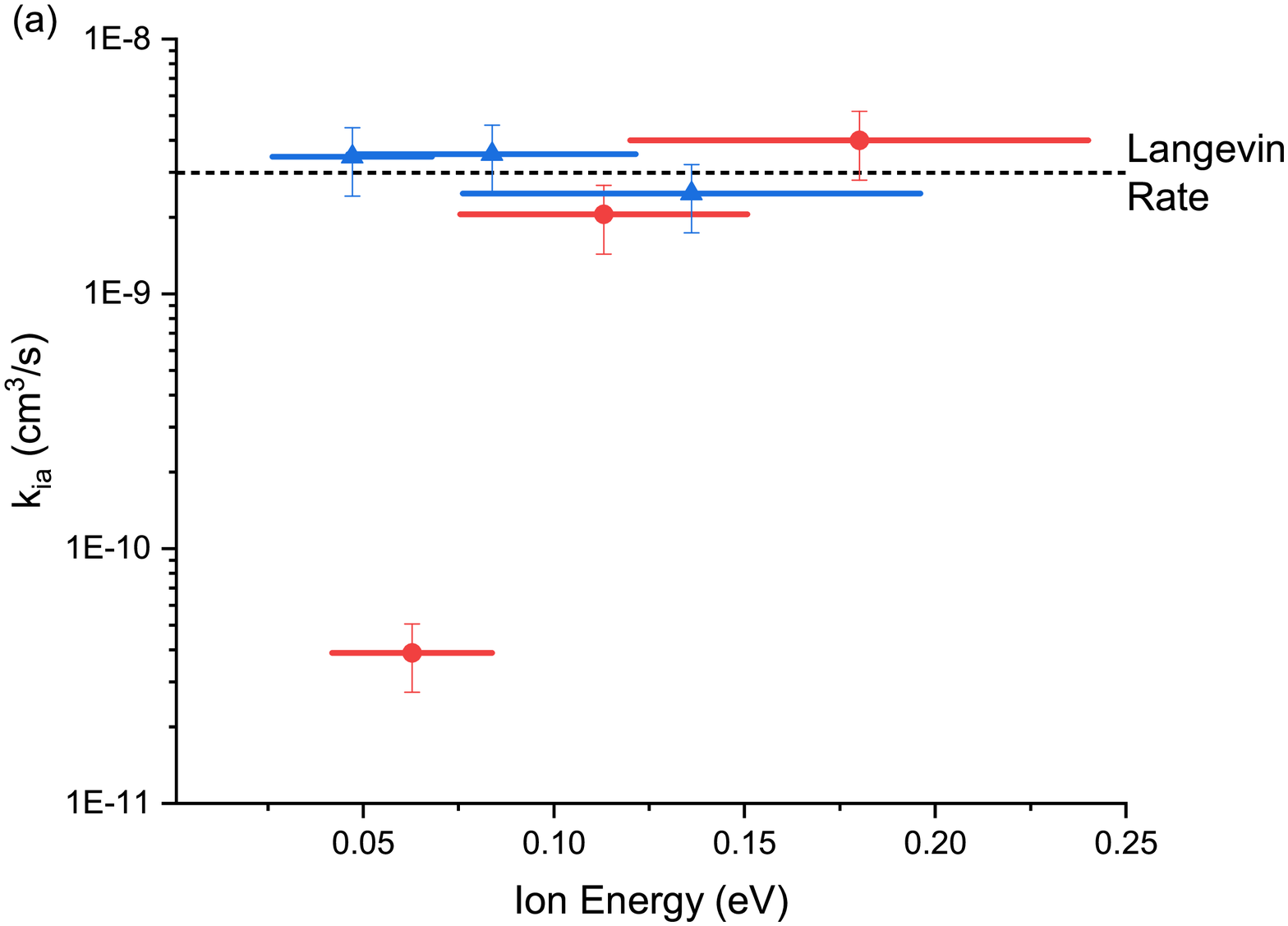}
\includegraphics[width=\linewidth]{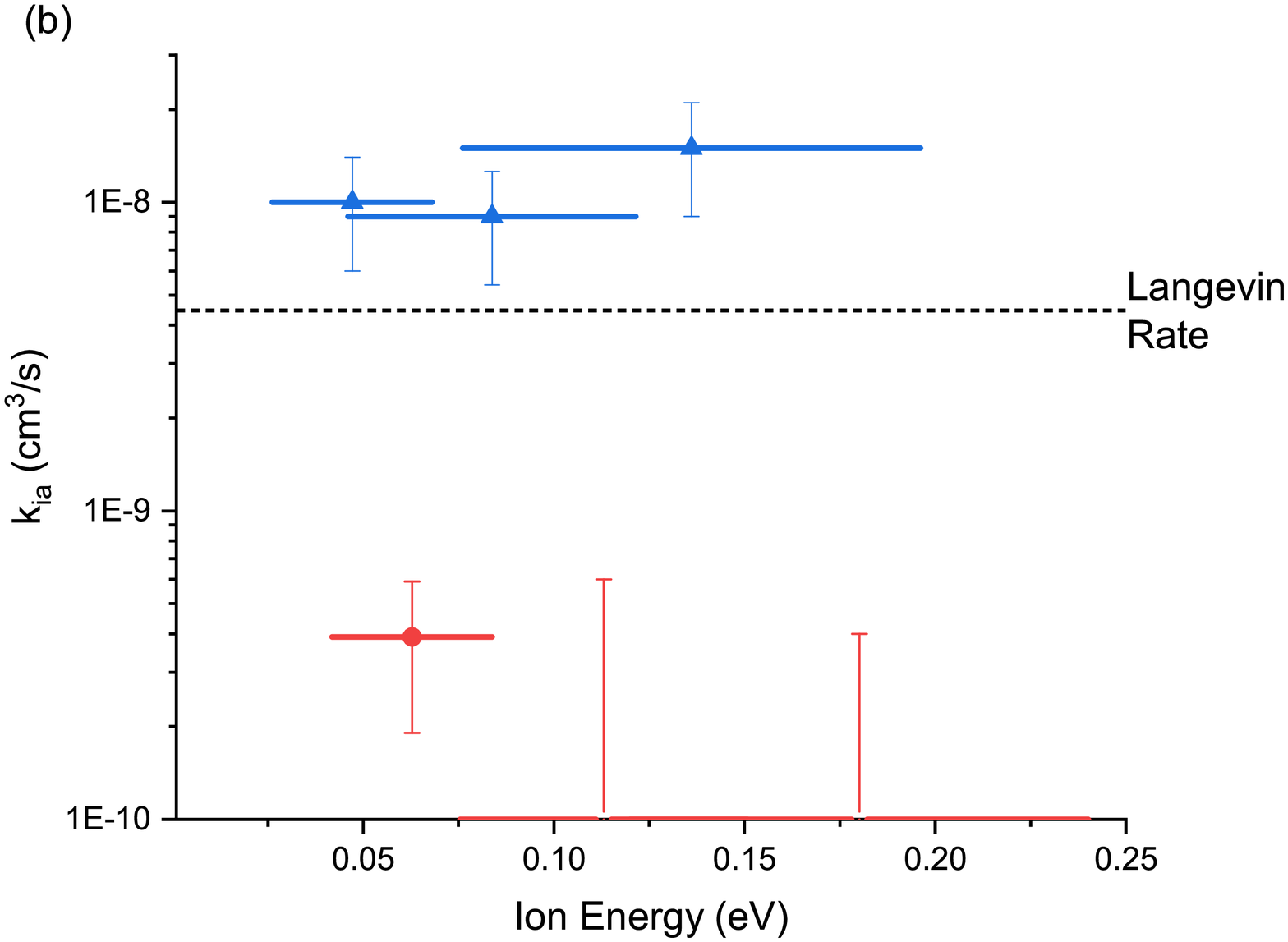}
\caption{\label{kvt} (Color online) The charge-exchange rate $k_\mathrm{ia}$ is shown as a function of ion collision energy for the neutral Na ground S-state (a) and excited P-state (b). In each plot, changing the state of the \ce{Ca+} results in very different rates for S-state (red circles) and D-state (blue triangles) ions. For reference, the classical energy-independent Langevin rate is shown as a dashed line. The reaction between $ \text{Na}[\text{P}]$ and $\text{Ca}^+[\text{D}]$ is above the Langevin limit (see the text for details). There is an increase in the $\text{Na}[\text{S}]+\text{Ca}^+[\text{S}]$ reaction channel as a function of temperature, while the other reaction channels remain fairly temperature insensitive. Two points in plot (b) were not determined to be significantly different from zero, so only the points' error bars are plotted. The horizontal error bars are bolded to represent the spread of energy of the ion distribution, as determined by the micromotion energy.}
\end{figure}

Measurements were performed using both S- and D-state \ce{Ca+} ions, which can be seen in Fig.~\ref{ratefe}. From these fits to Eq.~\eqref{eq:linfit}, we extrapolate individual $k_\mathrm{ia}$ rates for the entrance channels $\text{Na}[\text{S}]+\text{Ca}^+[\text{S~or~D}]$ (Fig.~\ref{kvt}a) and $\text{Na}[\text{P}]+\text{Ca}^+[\text{S~or~D}]$ (Fig.~\ref{kvt}b). In the D-state measurement, we must use an effective $(0.45)f_e$, since the MOT laser is being chopped at a 45\% duty cycle. 

The error for each value of $\gamma_\mathrm{ia}/\langle n\rangle$ was calculated by combining the fit-error from $\gamma_\mathrm{ia}$ with the propagated uncertainty from $\langle n\rangle$. Since $\langle n\rangle$ depends on $N_a$ and the ion and atom-cloud sizes, we propagate all of those errors into the uncertainty in $\langle n\rangle$.

A classical upper bound, $k_L$, on the charge-exchange rate is predicted by the Langevin scattering model \cite{grier_observation_2009}, as
\begin{equation}
k_L=2\pi \sqrt{\frac{\alpha e^2}{\mu(4\pi\epsilon_0)^2}}, \label{eq:langevin}
\end{equation}
which is collision-energy independent. The singly-ionized \ce{Ca+} has a net charge \mbox{$e\approx1.6\times10^{-19}$ C}. Here,  $\alpha/4\pi\epsilon_0 =24.11\times 10^{-30}\;\text{m}^3$, and $\alpha^*/4\pi\epsilon_0 =53.4\times 10^{-30}\;\text{m}^3$ are the atomic polarizabilities for the Na[S] and Na[P] states, respectively \cite{ekstrom_measurement_1995}. The reduced mass of the colliding system is $\mu =2.44\times 10^{-26}\;\text{kg}$. Using these numbers in Eq.~\eqref{eq:langevin} yields $k_L^{(g)}\approx 2.995\times 10^{-9}\;\text{cm}^3/\text{s}$ and $k_L^{(e)}\approx 4.457\times 10^{-9}\;\text{cm}^3/\text{s}$ for the \ce{Ca+} + Na[3S] and \ce{Ca+} + Na[3P], respectively. The classical Langevin model assumes unit efficiency for electron-capture within a critical radius. Thus, its rate coefficient is a classical upper bound for a long-range $V\propto -1/R^4$ potential. Except for the $\text{Na}[\text{P}]+\text{Ca}^+[\text{D}]$ entrance channel, the rates shown in Fig.~\ref{kvt} agree with the Langevin limit. Hall {\em et al.} found a charge-exchange rate in the $\text{Rb}+\text{Ca}^+$ system which was a factor of 4 {\it over} the classical Langevin prediction. They attribute this enhancement to the quadrupole interaction ($V\propto -1/R^3$ \cite{krych_sympathetic_2011}) with the P-state of Rb as well as near-resonance of the entrance and exit states \cite{hall_millikelvin_2012}. The increase we observe over Langevin in the $\text{Na}[\text{P}]+\text{Ca}^+[\text{D}]$ entrance channel may also be due to an additional quadrupole term in the long range interaction potential.

Accounting for quantum effects can result in a $k_\mathrm{ia}$ temperature dependence. Typically, the rate coefficient will increase as temperature increases, and can exhibit resonance-like behavior for very low energies \cite{cote_classical_2000}. However, most often this temperature dependence is weak (e.g., semiclassically $\sim T^{1/6}$ \cite{cote_classical_2000}). We find that the reaction rate for the $\text{Na}[\text{S}]+\text{Ca}^+[\text{S}]$ entrance channel varies by two orders of magnitude over a factor of three change in energy, as seen in Fig.~\ref{kvt}a. This is atypical, providing further evidence that the reaction channel has some collision-energy threshold ($\approx 0.05\;\text{eV}$).

For the $\text{Na}[\text{S}]+\text{Ca}^+[\text{S}]$ entrance channel, there exists only one lower-energy exit channel, $\text{Ca}[\text{S}]+\text{Na}^+[\text{S}]$. Charge-exchange via radiative charge transfer into this exit state is the only possible exothermic process. However, exothermic radiative charge transfer has been studied theoretically \cite{makarov_radiative_2003}, and there are two results from that study which suggest that this pathway is unlikely to be the dominant charge-exchange process. First, this exothermic process is predicted to be very slow, with a rate $k_\mathrm{ia}\sim 10^{-16} \;\text{cm}^3/\text{s}$, as compared to our experimental results. Second, an activation barrier on this entrance channel is not theoretically predicted to exist, yet we observe a collision-energy threshold experimentally. Therefore, we predict that the dominant process is endothermic. The closest endothermic exit channel has a separated atom limit of $\text{Ca}[4s4p \;^3P^0]+\text{Na}^+[2p^6 \;^1S]$, which has an energy $\approx .914\;\text{eV}$ above the entrance channel of $\text{Na}[\text{S}]+\text{Ca}^+[\text{S}]$ \cite{NIST_ASD}. The energy-gap between the separated-atom limits is much larger than the trapped-ion energies, so the reaction probably does not occur at large internuclear separation.

Alternatively, we predict that the molecular curve of the exit channel dips below its asymptote at small internuclear separation, allowing for charge exchange via non-radiative molecular association into the exit channel. However, we did not find any evidence of co-trapped \ce{NaCa+}, as discussed at the end of Sec.~\ref{subsec:decay}. Therefore, we also predict that the newly-formed molecular ions quickly dissociate via a secondary photodissociation process that uses the MOT beams' \mbox{589 nm} radiation. There may be several dissociative exit channels that could simultaneously be populated by the 589 nm photon. Finally, the dissociated $\text{Ca} + \text{Na}^+$ products are lost from the trap via MSRQ. Technically, the experiment measures the combined rate from all these processes.

The decay curves in Fig.~\ref{decays} suggest that the reaction with \ce{Ca+}[D] should also have a collision-energy threshold. However, the D-state $k_\mathrm{ia}$ data appear fairly temperature insensitive in Fig.~\ref{kvt}, suggesting that the D-state reaction threshold is lower than our range of measured energies. The \ce{Ca+}[D] reactions could also be endothermic or be a result of non-radiative charge transfer via an avoided crossing. However, we surmise that the dominant process is likely to be an exothermic reaction with a barrier in the entrance channel's molecular curve, \ce{Na}[S and/or P] + \ce{Ca+}[D]. This hypothesis is supported by our evidence of an increased elastic-scattering rate with the \ce{Ca+} D-state, which could also be caused by the same entrance-channel barrier [as discussed in Sec.~\ref{subsec:temp}].

\section{Conclusions}
\label{sec:concl}
In this work, we have measured charge-exchange reaction rates associated with four individual entrance channels of the Na + \ce{Ca+} system. We measure a high rate (between $10^{-11}\text{ and }10^{-9}\;\text{cm}^3/\text{s}$) of charge-exchange on the \ce{Na}[S] + \ce{Ca+}[S] entrance channel, which shows a significant collision-energy dependence. We predict that this reaction is an endothermic channel not quantitatively calculated by the original theoretical study on this system\cite{makarov_radiative_2003}. The largest charge exchange rates were observed for the \mbox{\ce{Na}[S, P] + \ce{Ca+}[D] }($\sim 10^{-8}$ cm$^3$/s). In the \mbox{$\text{Na}[\text{P}]+\text{Ca}^+[\text{D}]$} entrance channel, a rate higher than the classical Langevin prediction was observed, which is probably due to an additional ion and atomic-quadrupole interaction \cite{hall_millikelvin_2012}, not accounted for within the classical Langevin model.

In the future, we would like to laser cool the \ce{Ca+} and measure the mixture of $\text{Ca}^+[\text{3S,3P}]$ on $\text{Na} [\text{3S,3P}]$. By laser cooling \ce{Ca+}, we could also re-measure the \ce{Ca+} S-state reaction, but at much lower collision energy. Lastly, by loading the LPT to saturation with \ce{Ca+}, the loss rate from the MOT (rather than the LPT) could be used to determine the total $k_{ia}$ from charge-exchange and elastic scattering \cite{goodman_measurement_2015}. With the total rate-coefficient and having separately measured the charge-exchange rate here, the elastic rate could be determined and compared with theory \cite{makarov_radiative_2003}.

\section{Acknowledgements}

We would like to acknowledge support from the NSF under Grant No. PHY-1307874. We also thank Sam Entner for his help with the MOT temperature and $f_e$ measurements.



%

\end{document}